\documentclass[10pt,aps,prd,twocolumn,floats,floatfix,showpacs,superscriptaddress,nofootinbib,longbibliography]{revtex4-1}
\usepackage[utf8]{inputenc}               		

\usepackage{graphicx,mathtools,amssymb,amsmath,amsthm,amsfonts,epsfig,epsf,bm}
\usepackage[outdir=./]{epstopdf}
\usepackage[usenames]{color}		
\usepackage{tensor}	
\usepackage{csquotes}
\usepackage{mathrsfs}
\usepackage{autobreak}
\usepackage[caption=false]{subfig}
\usepackage{enumerate}
\usepackage[T1]{fontenc}

\DeclareSymbolFont{TOneChars}{T1}{\familydefault}{m}{it}
\SetSymbolFont{TOneChars}{bold}{T1}{\familydefault}{bx}{it} 
\DeclareMathSymbol{\mathdh}{\mathord}{TOneChars}{"F0}

\newcommand{\GB}{\mathcal{G}}
\newcommand{\nn}{\mathbf{n}}

\usepackage[dvipsnames]{xcolor}

\usepackage[linktocpage]{hyperref}
\hypersetup{hidelinks}
\hypersetup{pdfstartview=}
\hypersetup{
    colorlinks=true,
    linkcolor=blue,
    filecolor=magenta,      
    urlcolor=BrickRed,
    citecolor=BrickRed,
}

\usepackage[capitalise]{cleveref}

\begin{document}

\title{Frequency contamination from new fundamental fields in black hole ringdowns}

\author{\textbf{Jacopo Lestingi}}
\affiliation{Nottingham Centre of Gravity \& School of Mathematical Sciences, University of Nottingham,
University Park, Nottingham NG7 2RD, United Kingdom}
\author{\textbf{Giovanni D'Addario}}
\affiliation{Nottingham Centre of Gravity \&  School of Mathematical Sciences, University of Nottingham,
University Park, Nottingham NG7 2RD, United Kingdom}
\affiliation{School of Physics and Astronomy, University of Nottingham,
University Park, Nottingham NG7 2RD, United Kingdom}
\author{\textbf{Thomas P. Sotiriou}}
\affiliation{Nottingham Centre of Gravity \&  School of Mathematical Sciences, University of Nottingham,
University Park, Nottingham NG7 2RD, United Kingdom}
\affiliation{School of Physics and Astronomy, University of Nottingham,
University Park, Nottingham NG7 2RD, United Kingdom}

\begin{abstract}

We revisit the modelling of black hole ringdown beyond General Relativity (GR), emphasizing the limitations of approaches that rely solely on shifted quasinormal mode (QNM) frequencies. Starting from modified Teukolsky equations in such scenarios, we classify the distinct types of deviations that can arise --- those shifting QNM frequencies, and those introducing additional frequencies associated with extra fields. We then construct the most general ansatz for metric perturbations in this context and discuss its implications for QNM modelling and theory-agnostic tests of GR using gravitational wave data.

\end{abstract}

\maketitle

\section{Introduction}

The final oscillations of a black hole (BH) settling to its quiescent state, known as the ringdown, can be modelled, at least in part, by a sum of quasinormal modes (QNMs) \cite{Kokkotas:1999bd,Nollert:1999ji,Berti:2009kk}. In General Relativity (GR) the frequencies of the modes depend on the mass and spin of the final BH. 
With the advent of gravitational wave (GW) observations \cite{LIGOScientific:2018mvr,LIGOScientific:2020ibl,KAGRA:2021vkt}, new tests of GR in the strong gravity regime can be performed, and the ringdown phase following a binary merger is one of them. 

Deviations from GR in the ringdown signal could appear if a new field that is part of the extension of GR or the Standard Model affects the structure or dynamics of BHs. Charges associated with such field are colloquially referred to as \textit{hair}. Although several \textit{no-hair} theorems exist, especially for scalar fields \cite{Hawking:1972qk,Bekenstein:1995un,Sotiriou:2011dz,Hui:2012qt,Herdeiro:2015waa,Sotiriou:2015pka},
there are also interesting exceptions \cite{Kanti:1995vq,Yunes:2011we,Babichev:2013cya,Sotiriou:2013qea,Sotiriou:2014pfa,Herdeiro:2014goa,Antoniou:2017hxj,Silva:2017uqg,Doneva:2017bvd,Dima:2020yac,Herdeiro:2020wei,Antoniou:2021zoy,Doneva:2022ewd}.

BH hair could conceivably affect the observed GWs, providing a smoking gun for the presence (or absence) of deviations from GR; this is a key objective of the \textit{black hole spectroscopy} programme \cite{Dreyer:2003bv,franchini2024testing}. Indeed,
in recent years, significant progress has been made in the understanding of BH perturbations beyond GR (bGR). Modified Teukolsky equations for a wide class of theories have been derived \cite{Li:2022pcy,Hussain:2022ins,Li:2023ulk,Cano:2023tmv} and applied to compute quasinormal frequencies \cite{Cano:2024ezp,Cano:2024jkd,Weller:2024qvo,Li:2025fci}.

An alternative to studying BH perturbations for a specific theory is to attempt to develop a \textit{theory-agnostic} parameterisation that captures generic bGR features.
Attempts in this direction  traditionally model deviations from GR by constructing a sum of QNMs with a frequency which is shifted with respect to the GR value, i.e.
\begin{equation}
    \omega_{bGR} = \omega_{GR} + \delta \omega \;,
\end{equation}
where $\delta \omega$ is a small shift in the frequency. The objective of such analyses is then to search for evidence of nonzero $\delta \omega$ in GW data \cite{LIGOScientific:2020tif,LIGOScientific:2021sio}.

Here we draw attention to and seek to clarify an important feature of bGR ringdowns: the presence of additional fields (or degrees of freedom) nonminimally coupled to gravity leads to a coupling of linear perturbations of the metric and the new fields. This in turn implies that it is insufficient to characterise the ringdown with a single set of shifted frequencies.
This linear coupling and the existence of solutions that contain the frequencies of the new field have both been  noted in numerical studies of QNMs of bGR BHs in specific scalar-tensor theories of gravity, see for example \cite{Molina:2010fb,Blazquez-Salcedo:2016enn}.
In this study, we will first argue that the coupling of linear perturbations around black hole spacetime is a generic feature for extra fields nonminimally coupled to gravity. We will then show analytically that the ringdown signal will include the characteristic QNM frequencies  of the extra fields, in addition to the shifted QNM frequencies of the metric.  We call this generic feature \textit{frequency contamination}.

Crucially, we show that under realistic conditions, the correction to the GR ringdown waveform coming from these contaminated gravitational modes is of the same order of magnitude as the correction due to QNM frequency shifts.

We use the failure of the Wald identity \cite{PhysRevLett.41.203} to hold in algebraically general spacetimes as a starting point to lay out the formal structure of a modified Teukolsky equation (which indeed matches the structure of the Teukolsky equation derived in detail in \cite{Li:2022pcy}), and discuss the various contributions to it. 
Analyzing this equation, we derive the most general form of the ansatz for metric perturbations which should be employed in ringdown tests of GR. We also show that our result agrees with the ansatz recently proposed in \cite{Crescimbeni:2024sam} in the context of theory-agnostic QNM searches. According to our analysis,  the dynamics of the bGR BH spacetime is captured by a hierarchical system, provided that the extra field is weakly coupled to gravity. The characteristic modes of the extra field get imprinted on the gravitational radiation. From the perspective of spectroscopy, this amounts to a contamination of the ringdown signal by the characteristic frequencies of the new field in the background spacetime.

The paper is structured as follows. In \cref{sec: setup} we describe the classes of bGR theories that we are considering, before discussing the master equations in such theories in \cref{sec:Teukolsky}.
Finally, in \cref{sec: ringdown ansatz} we illustrate how to construct a general ansatz for modelling bGR ringdown, before discussing the physical implications of the ansatz in \cref{sec:Discussions}.

\section{Setup} \label{sec: setup}

We consider theories that can be parameterised as
\begin{equation}
    \mathcal{L} = \mathcal{L}_{\mathrm{GR}} + \ell^p \mathcal{L}_{\mathrm{bGR}} +  \mathcal{L}_{\mathrm{field}}  +  \mathcal{L}_{\mathrm{matter}}\; ,
    \label{eq:ModifiedTheories}
\end{equation}
where $\mathcal{L}_{\mathrm{GR}}$ is the Einstein-Hilbert Lagrangian, $\mathcal{L}_{\mathrm{bGR}}$ contains all bGR interactions, such as higher curvature terms or nonminimal couplings with extra, dynamical fields, while $\mathcal{L}_{\mathrm{field}}$ is the Lagrangian for these dynamical fields. Note that in the case of higher derivative gravity, $\mathcal{L}_{\text{field}}=0$. Working in geometric units with $c = G = 1$, we take $\ell$ to have dimensions of a length, while $p$ is an integer determined by the requirement that $\ell^p \mathcal{L}_{\mathrm{bGR}}$ has the same dimensions as $\mathcal{L}_{\mathrm{GR}}$. Finally, we neglect $\mathcal{L}_{\mathrm{matter}}$, which describes any further matter fields.

Action \eqref{eq:ModifiedTheories} is general enough to describe any extra field that is nonminimally coupled to gravity. For illustrative purposes, and to provide a concrete example, we will take this extra field to be a scalar, $\Phi$. We stress that our results are valid for fields of different helicities as well, and hence $\Phi$ should be seen as a placeholder for any field. Examples of theories with an extra (pseudo)scalar which exhibit BH hair are dynamical-Chern-Simons (dCS) \cite{Alexander:2009tp,Yunes:2011we}  and scalar-Gauss-Bonnet (sGB) gravity \cite{Kanti:1995vq,Yunes:2011we,Sotiriou:2013qea,Sotiriou:2014pfa,Antoniou:2017hxj,Silva:2017uqg,Doneva:2017bvd,Dima:2020yac,Herdeiro:2020wei,Antoniou:2021zoy,Doneva:2022ewd}.
In our parameterization of bGR theories, the latter can be written as $\mathcal{L}^{\mathrm{sGB}} = \mathcal{L}_{\mathrm{GR}} + \alpha_{\mathrm{sGB}} \mathcal{L}_{\mathrm{bGR}}^{\mathrm{sGB}} + \mathcal{L}_{\mathrm{field}}^{\mathrm{sGB}}$, with $\alpha_{\mathrm{sGB}} = \ell_{\mathrm{sGB}}^2$ and
\begin{align}
    \mathcal{L}_{\mathrm{GR}} & = \frac{R}{16 \pi}\\
    \mathcal{L}_{\mathrm{bGR}}^{\mathrm{sGB}} & = \frac{f(\Phi)\GB}{16 \pi}\\
    \mathcal{L}_{\mathrm{field}}^{\mathrm{sGB}} & = -\frac{1}{32 \pi} \nabla^\mu \Phi \nabla_\mu \Phi \; ,
\end{align}
where $\GB = R^{\mu\nu\lambda\kappa}R_{\mu\nu\lambda\kappa} - 4R^{\mu\nu}R_{\mu\nu} + R^2$ is the Gauss-Bonnet invariant, $\Phi$ is a scalar field and $f(\Phi)$ describes the interaction between the scalar and the Gauss-Bonnet.

The field equation for a scalar field $\Phi$ with a canonical kinetic term and a potential $V(\Phi)$ is:
\begin{equation}
\label{eq:EOM Extra field}
    \Box_{g} \Phi  = -\ell^p \left( \Sigma_{A}[g, \Phi] + \Sigma_{B}[g] \right)-  V'(\Phi),
\end{equation}
\begin{equation}
\frac{16 \pi}{\sqrt{-g}}\frac{\delta \mathcal{L}_{\mathrm{bGR}}}{\delta \Phi} = \Sigma_{A}[g, \Phi] + \Sigma_{B}[g]\; .
\end{equation}
Note that source terms such as $\Sigma_B[g]$ with no dependence on $\Phi$ can arise in theories in which $\Phi$ only appears linearly in the coupling to gravity. A relevant example is linear Gauss-Bonnet  gravity \cite{Sotiriou:2013qea,Sotiriou:2014pfa}. By separating this source term explicitly, we have that $\Sigma_A[g,\Phi]$ depends at least linearly on $\Phi$. In the rest of the paper we set $V(\Phi)=0$.

The equations for the metric can be viewed as Einstein's equations with  additional source terms:
\begin{equation}
\label{General Einstein equation}
    G_{\mu\nu}[g] = T^\text{field}_{\mu\nu}[g,\Phi] + \ell^p T_{\mu\nu}^{\mathrm{bGR}}[g, \Phi] \; .
\end{equation}
Note that, if one allows for the presence of higher derivative gravity terms in scalar-tensor theories of gravity,  then $T_{\mu\nu}^{\mathrm{bGR}}[g, \Phi]$ will in general feature a scalar-independent term $T_{\mu\nu}^{\mathrm{bGR}}[g]$\footnote{It is worth clarifying that modifications of GR arising from higher order curvature invariants generically introduce additional degrees of freedom and can be brought in a form compatible with action \eqref{eq:ModifiedTheories} after field redefinitions. In cases where such degrees of freedom are considered spurious and removed in some way (e.g. perturbative expansions in the context of Effective Field Theories), they would clearly not affect ringdowns.}.

\subsection{Perturbative formalism} \label{sec: perturbative formalism}

In the context of modelling BHs in bGR theories, it is common to work perturbatively in the dimensionless quantity $\ell^p/M^p$, where $M$ is the mass of the BH, under the assumption that corrections to GR are small. Indeed, our approach and action \eqref{eq:ModifiedTheories} are inspired by applications of this perturbative treatment to extreme mass ratio inspirals \cite{Maselli:2020zgv,Maselli:2021men,Barsanti:2022vvl,Spiers:2023cva,Speri:2024qak} and massive BHs \cite{DAddario:2023erc}, as well as in the context of the derivation of the modified Teukolsky equation \cite{Li:2022pcy}. Here we adopt the convention of \cite{Li:2022pcy}, where the dimensionless parameter
\begin{equation}
 \zeta = \left(\frac{\ell^p}{M^p} \right)^2
\end{equation} 
is defined such that at first order it captures the leading order bGR corrections to the metric (rather than the leading order correction to the scalar). To illustrate this, we consider the scalar field equation \eqref{eq:EOM Extra field}: without the source terms, this equation is covered by a no-hair theorem \cite{Sotiriou:2011dz}, which implies that, to leading order in $\ell/M$,  $\Phi \sim \ell^p/M^p$. Hence the leading order corrections to the metric, due to the backreaction of the scalar, enter at order $(\ell^p/M^p)^2$.

With this consideration in mind, we can rewrite \eqref{General Einstein equation} as:
\begin{equation}
\label{eq: Einstein Equation}
    G_{\mu\nu}[g(\zeta)] = T_{\mu\nu}[g(\zeta), \Phi],
\end{equation}
where for convenience we have collected the stress-energy tensor of the scalar, $T^{\text{field}}_{\mu\nu}$, with the contributions from the bGR couplings of the scalar to curvature into a single source, $T_{\mu\nu}$ as, to leading order, they both appear at $O(\zeta)$. This is because $T_{\mu \nu}^{\text{field}}$ is quadratic in $\Phi \sim \zeta^{1/2}$, and, likewise, $\zeta^{1/2}T_{\mu\nu}^{\mathrm{bGR}}[g(\zeta), \Phi] \sim \zeta$. Likewise, we can also rewrite \eqref{eq:EOM Extra field} as:
\begin{equation}
    \label{eq:EOM scalar field}
    \Box_{g(\zeta)} \Phi  = - \zeta^{1/2} \left( \Sigma_{A}[g(\zeta), \Phi] + \Sigma_{B}[g(\zeta)] \right),
\end{equation}
where, for convenience, we use again $\Sigma_{A,B}$ even though their dimensions are rescaled with respect to \eqref{eq:EOM Extra field}.

In the above we have assumed that $p>0$, which implies that bGR corrections are suppressed by some characteristic mass scale (the inverse of $\ell$) in particle physics units. The $p=0$ case is somewhat special for scalars, as it corresponds to a coupling (without derivatives) to $R$. This can be removed by a conformal transformation and is also known to not lead to BH hair \cite{Hawking:1972qk,Sotiriou:2011dz}. More broadly however, and for any new field, if $p=0$, the $\zeta$ can be identified as the dimensionless coupling that appears in the action. The perturbative treatment we lay out below will continue to apply provided that this coupling is sufficiently small and the background solution is continuously connected to GR as $\zeta\to 0$.

A spacetime described by equation \eqref{eq: Einstein Equation}, will in general not feature all the symmetries of a Kerr spacetime. In particular, it may not be of Petrov type D \cite{Petrov:2000bs}, key to the derivation of a Teukolsky equation \cite{Teukolsky:1973ha} that greatly simplifies the study of gravitational radiation propagating in said spacetime. Indeed, in many modified theories of gravity, such as sGB or dCS, the spacetime is said to be algebraically general, or Petrov type I \cite{Owen:2021eez}.

As we are interested in dynamical perturbations to BHs within a bGR setting, we can define a perturbative formalism with two expansion parameters, the first of which is $\zeta$. We introduce a second parameter $\epsilon$ to indicate GW/dynamical perturbations.
As we only consider dynamical perturbations to order $\epsilon$ (i.e.  we are not considering nonlinear dynamical perturbations, such as quadratic QNMs \cite{Campanelli:1998jv}), we use a single superscript $(n)$ to denote unperturbed $ (n = 0)$ and perturbed $(n=1)$ quantities in $\zeta$, and employ separate letters to indicate dynamical $O(\epsilon)$ and nondynamical $O(\epsilon^0)$ perturbations. For example, a metric and a scalar field can be expanded respectively as:
\begin{align}
     g_{\mu\nu}  & =  g^{(0)}_{\mu\nu} + \epsilon h^{(0)}_{\mu\nu} + \zeta g^{(1)}_{\mu\nu} + \zeta\epsilon h^{(1)}_{\mu\nu} + \dots \\
     \Phi & = \Phi^{(0)} + \epsilon \varphi^{(0)} + \zeta^{1/2} \Phi^{(1)} +  \zeta^{1/2} \epsilon \varphi^{(1)} + \dots 
\end{align}
As stated above, $\Phi \sim \zeta^{1/2}$, which is equivalent to assuming $\Phi^{(0)}=\varphi^{(0)}=0$.

\subsection{Linearized field equations}

Consider the field equation for $\Phi$, \eqref{eq:EOM scalar field}. To derive an equation for the leading term $\Phi^{(1)}$, we expand \eqref{eq:EOM scalar field} in $\zeta$, neglecting $O(\zeta)$ and higher terms. The $O(\zeta^{1/2})$ equation reads
\begin{equation}
\label{eq: Phi1 Equation}
    \Box_{g^{(0)}} \Phi^{(1)} = -\Sigma_B^{(0)}[g^{(0)}] \; .
\end{equation}
The operator $\Box_{g^{(0)}}$, which is defined in the Kerr background, coincides with the spin-0 Teukolsky operator, $\hat{\mathcal{O}}^{(0)}$: hence,
\begin{equation}
\label{eq: stationary scalar}
    \hat{\mathcal{O}}^{(0)} \Phi^{(1)} = -\Sigma_B^{(0)}[g^{(0)}] \; .
\end{equation}
Henceforth, we employ hats to denote operators, frequencies, and functions associated to the scalar field $\Phi$.
As we are assuming that the background is the Kerr metric, the right-hand side of \eqref{eq: stationary scalar} is stationary, and therefore $\Phi^{(1)}$ represents a stationary correction, which is consistent with our perturbative formalism. Indeed, the homogeneous solution to this equation, which will be nonstationary, will be part of the dynamical perturbation $\varphi^{(1)}$. To see this, we now perturb the original scalar equation \eqref{eq:EOM scalar field} in $\epsilon$. At $O(\epsilon)$, we find
\begin{align}
\label{eq:Scalar Field EOM Order epsilon}
    \Box_{h(\zeta)} \Phi + \Box_{g(\zeta)} \varphi  = & -\zeta^{1/2} \Big( \sigma_{A,h}[h(\zeta), \Phi] + \sigma_{A,\varphi}[g(\zeta), \varphi] \notag \\
    &+ \sigma_B[h(\zeta)] \Big) \; .
\end{align}
Here $\sigma_B[h(\zeta)]$ arises from the linearization in $\epsilon$ of $\Sigma_B[g(\zeta)]$. The $\epsilon$-linearization of $\Sigma_A[g(\zeta),\Phi]$ is more subtle; as both $g(\zeta)$ and $\Phi$ can be expanded in $\epsilon$, at $O(\epsilon)$ we find two contributions that we call $\sigma_{A,h}[h(\zeta), \Phi]$ and $\sigma_{A,\varphi}[g(\zeta), \varphi]$.  
We then perturb \eqref{eq:Scalar Field EOM Order epsilon} in $\zeta$; at $O(\zeta^{1/2}\epsilon)$, we obtain
\begin{align}
    \Box_{h^{(0)}} \Phi^{(1)} + \Box_{g^{(0)}} \varphi^{(1)}  = & \sigma_{A,h}^{{(0)}}[h^{(0)}, \Phi^{(0)}] \notag \\
    & + \sigma_{A,\varphi}^{{(0)}}[g^{(0)}, \varphi^{(0)}] + \sigma_B^{(0)}[h^{(0)}] \; .
\end{align}
The first two source terms vanish as $\Phi^{(0)}  =\varphi^{(0)} = 0$, and $\sigma_A$ is at least linear in $\Phi$, as was the case for $\Sigma_A$, by construction. Hence, this equation becomes
\begin{equation}
\label{eq: Varphi1 Eqn}
     \hat{\mathcal{O}}^{(0)} \varphi^{(1)} = - \Box_{h^{(0)}} \Phi^{(1)}   + \sigma_B^{(0)}[h^{(0)}] =  -\sigma^{(0)}[\Phi^{(1)}, h^{(0)}]\; .
\end{equation}
We therefore see that the homogeneous solution of the spin-0 Teukolsky equation is included in $\varphi^{(1)}$. In our notation, $\sigma_B[g]$ is only nonzero for theories including a linear coupling of the scalar to gravity, such as sGB with $f(\Phi) = \Phi + \dots$ .

We note that to compute $\varphi^{(1)}$ we need both $\Phi^{(1)}$, that can be computed via \eqref{eq: stationary scalar}, but also $h^{(0)}$, the gravitational perturbation of the Kerr background. As usual, $h^{(0)}$ satisfies the linearized Einstein equation on the Kerr background, namely:
\begin{equation}
    \mathcal{E}_{\mu \nu}^{(0)}[h^{(0)}] = 0,
\end{equation}
where $\mathcal{E}^{(0)}_{\mu\nu}$ is the linearized Einstein tensor on a Kerr background, and reads:
\begin{align}
\label{linearized Einstein operator}
    \mathcal{E}^{(0)}_{\mu \nu}[h^{(0)}] & \equiv \frac{1}{2} \Big[ - \nabla^{\rho} \nabla_{\rho} h^{(0)}_{\mu \nu} - \nabla_{\mu} \nabla_{\nu} (g^{(0)\rho \sigma}h_{\rho \sigma}^{(0)})  
\nonumber \\ &
+ 2 \nabla^{\rho} \nabla_{(\mu} h^{(0)}_{\nu)\rho} + g^{(0)}_{\mu \nu} (\nabla^{\rho} \nabla_{\rho} h^{(0)} - \nabla^{\rho} \nabla^{\sigma} h^{(0)}_{\rho \sigma}) \Big],
\end{align}
where the covariant derivative is defined using the background metric $g^{(0)}_{\mu \nu}$.

We now proceed to derive the linearized bGR Einstein equation at order $O(\zeta \epsilon)$, starting from \eqref{eq: Einstein Equation}
\begin{equation}
\label{eq: Einstein Equation No Indices}
    G[g(\zeta)] = T[ g(\zeta), \Phi] \; ,
\end{equation}
where we have suppressed the spacetime indices for convenience.
Note that, according to our notation, a bGR BH solution of this equation reads $g^{(0)}_{\mu\nu}+\zeta g^{(1)}_{\mu\nu}$, up to first order in $\zeta$.
Expanding \eqref{eq: Einstein Equation No Indices} in $\epsilon$, we find at $O(\epsilon)$:
\begin{equation}
    \mathcal{E}[h(\zeta)] = \tau_\varphi [ g(\zeta),\varphi ] + \tau_h [  h(\zeta), \Phi] \;,
\end{equation}
where we have employed the same linearization notation for  $\tau_\varphi$ and $\tau_h$ as we did with $\sigma_{A,\varphi}$ and $\sigma_{A,h}$.
Subsequently, we can expand in $\zeta$:
\begin{align}
    \mathcal{E}^{(0)}[ h^{(1)}] + \mathcal{E}^{(1)}[ h^{(0)}] &= \tau_{\varphi}^{(0)} [ g^{(0)}, \varphi^{(1)}] + \tau_{\varphi}^{(0)} [ g^{(1)}, \varphi^{(0)} ] \notag \\
    & + \tau_{\varphi}^{(1)} [ g^{(0)}, \varphi^{(0)} ] +  \tau_{h}^{(0)} [ h^{(0)}, \Phi^{(1)}  ] \notag\\
& + \tau_{h}^{(0)} [ h^{(1)}, \Phi^{(0)}  ] + \tau^{(1)}_h [ h^{(0)}, \Phi^{(0)} ] \; .
\end{align}
By construction, $T[g(\zeta),\Phi]$ always depends at least linearly on $\Phi$. This dependence on the scalar field is inherited by $\tau_{\varphi}$ and $\tau_h$. Hence: $\tau_{\varphi}^{(0)} [ g^{(1)}, \varphi^{(0)} ] = \tau_{\varphi}^{(1)} [ g^{(0)}, \varphi^{(0)} ] = \tau_{h}^{(0)} [ h^{(1)}, \Phi^{(0)}  ] = \tau^{(1)}_h [ h^{(0)}, \Phi^{(0)} ] = 0$. We are therefore left with
\begin{equation}
\label{eq: Linearised Einstein equation bGR}
    \mathcal{E}^{(0)}[ h^{(1)}] + \mathcal{E}^{(1)}[ h^{(0)}] = \tau_{\varphi}^{(0)} [ g^{(0)}, \varphi^{(1)}]+  \tau_{h}^{(0)} [ h^{(0)}, \Phi^{(1)}  ]  \; ,
\end{equation}
which is the $O(\zeta)$ term of the bGR linearized Einstein equation.

\section{Structure of the modified Teukolsky equation\label{sec:Teukolsky}}

In the following, $\{ l^{\mu},n^{\mu},m^{\mu},m^{*\mu}\}$ is the standard tetrad of the Kerr spacetime. The derivation of the (modified) Teukolsky equation relies on the Newman-Penrose or Geroch-Held-Penrose (GHP) formalisms \cite{Newman:1961qr,Geroch:1973am}; we review both formalisms in Appendix \ref{sec: appendix GHP}. Of particular importance are the Weyl scalars $\Psi_{0,1,2,3,4}^{(0)}$, which can be obtained by projecting the Weyl tensor $C_{\alpha \beta \gamma \delta}^{(0)}$ onto combinations of the tetrad components. The Weyl scalars $\Psi_0^{(0)} \equiv -C_{\alpha \beta \gamma \delta}^{(0)} l^{\alpha}m^{\beta}l^{\gamma}m^{\delta}$ , $\Psi_4^{(0)} \equiv -C_{\alpha \beta \gamma \delta}^{(0)} n^{\alpha}m^{*\beta}n^{\gamma}m^{*\delta}$ are especially relevant, as they correspond to fields of spin weight $s=2$ and $s-2$ respectively and contain all of the information about gravitational perturbations of the Kerr metric. Here, we have focused on the Weyl scalars of the Kerr solution; however, they can be defined analogously in a generic spacetime.

Following our notation for perturbations in $\zeta$ and $\epsilon$, we can define the dynamical perturbations $\psi$ of $\Psi_0$ as
\begin{equation}
     \Psi_0  = \Psi^{(0)}_0 + \epsilon \psi^{(0)} + \zeta \Psi^{(1)} + \zeta\epsilon \psi^{(1)} + \dots. 
\end{equation}
It is well known that to study linear perturbations of a Kerr background (or, more generally, vacuum Petrov type D), the equation to study is the Teukolsky equation: $\mathcal{O}^{(0)}_s\psi^{(0)}_s=0$ \cite{Teukolsky:1973ha}, where $s=\pm 2$ is the spin weight of the field $\psi_s$, and
\begin{align}
\psi^{(0)} & \equiv \psi^{(0)}_{s=2} \; ,  \quad \tilde \psi^{(0)} \equiv \psi^{(0)}_{s=-2} \;, \\
    \mathcal{O}^{(0)} & \equiv \mathcal{O}^{(0)}_{s=2} =  g^{(0)\mu \nu} (\Theta_{\mu} + 4 B_{\mu})(\Theta_{\nu} + 4 B_{\nu}) \notag \\
   & \hspace{4.5 em} - 16 \Psi_2^{(0)},
   \label{eq: Teukolsky operator}\\ 
   \mathcal{O}^{\dagger(0)} & \equiv \mathcal{O}^{(0)}_{s=-2} = g^{(0)\mu \nu} (\Theta_{\mu} - 4 B_{\mu})(\Theta_{\nu} - 4 B_{\nu}) \notag \\
   & \hspace{5em} - 16 \Psi_2^{(0)}.
\end{align} 
Here, $\Theta$ is the GHP covariant derivative (see Appendix \ref{sec: appendix GHP}), defined with respect to the background Kerr metric, and $B^{\mu} = -(\rho n^{\mu}-\tau m^{*\mu})$, with $\rho, \tau$ spin coefficients of the Kerr spacetime. As explained in Appendix \ref{sec: appendix GHP}, $\mathcal{O}^{(0)}_{s=-2}$ is the formal adjoint of $\mathcal{O}^{(0)}$, and hence is defined as $\mathcal{O}^{\dagger(0)}$.

The derivation of decoupled equations for $\psi^{(0)},\tilde\psi^{(0)}$ is central to the existence of the Wald identity \cite{PhysRevLett.41.203}, an operatorial identity relating the linearized Einstein tensor with the Teukolsky operator $\mathcal{O}^{(0)},\mathcal{O}^{\dagger (0)}$.
The Wald identity in the Kerr spacetime reads:
\begin{equation}
    \mathcal{S}^{(0)}\mathcal{E}^{(0)} = \mathcal{O}^{(0)}\mathcal{T}^{(0)},
\end{equation}
which holds generally for vacuum Petrov type D spacetimes. Here, $\mathcal{E}^{(0)}$ is the linearized Einstein tensor, defined in \eqref{linearized Einstein operator},
$\mathcal{O}^{(0)}$ is the Teukolsky operator \eqref{eq: Teukolsky operator}, 
and $\mathcal{S}^{(0)}, \mathcal{T}^{(0)}$ are linear differential operators defined by \cite{Aksteiner:2014zyp}: 
\begin{align}
\label{S operator}
& \mathcal{S}^{(0)}u = Z^{\nu \rho \sigma \mu} (\Theta_{\mu} + 4B_{\mu}) \Theta_{\nu} u_{\rho \sigma},
\\ &
\label{T operator}
\mathcal{T}^{(0)}v = -\frac{1}{2} Z^{\nu \rho \sigma \mu} \Theta_{\mu} \Theta_{\nu} v_{\rho \sigma},
\end{align}
with $Z^{\mu \nu \rho \sigma} \equiv Z^{\mu \nu} Z^{ \rho \sigma}$, $Z^{\mu \nu} \equiv 2l^{[\mu} m^{\nu]}$, and  $u,v$ are generic twice covariant tensor fields. Specifically, $\mathcal{T}^{(0)}$ is the differential operator such that: $\psi^{(0)} = \mathcal{T}^{(0)}h^{(0)}$, which can be obtained by linearizing $\Psi_0^{(0)}$ in $\epsilon$.

If the Wald identity holds, the Teukolsky equation can be derived in a straightforward manner. Consider a solution $h^{(0)}$ to the linearized Einstein equation $\mathcal{E}^{(0)}[h^{(0)}]=0$. Using the Wald identity one has: $0 = \mathcal{S}^{(0)}\mathcal{E}^{(0)}[h^{(0)}] = \mathcal{O}^{(0)}\mathcal{T}^{(0)}h^{(0)}$. Then, using $\psi^{(0)} = \mathcal{T}^{(0)}h^{(0)}$, the $s=2$ Teukolsky equation $\mathcal{O}^{(0)}\psi^{(0)}=0$ follows. 

In this section, by generalizing this procedure, we show that it is possible to recover the same formal structure of the modified Teukolsky equation introduced in \cite{Li:2022pcy}, required to compute bGR corrections at order $\zeta$ to the $\epsilon-$perturbed $s=2$ Weyl scalar $\psi^{(0)}$. The description of the modified Teukolsky equation in the $s=-2$ case follows the same steps and is briefly discussed at the end of the section.

Consider now a spacetime with metric $g(\zeta)$. By linearizing the Einstein tensor $G(\zeta)$ and the Weyl scalar $\Psi_0(\zeta)$, one finds the differential operators $\mathcal{E}(\zeta)$ and $\mathcal{T}(\zeta)$. By construction, $\psi(\zeta) = \mathcal{T}(\zeta)h(\zeta)$. 
Furthermore, we can generalize the tensor  $\mathcal{S}^{(0)}$ to a spacetime with metric $g(\zeta)$ which is different from the Kerr metric by simply replacing $g^{(0)}$ with $g(\zeta)$, giving $\mathcal{S}(\zeta)$. Finally, the Teukolsky operator $\mathcal{O}^{(0)}$ becomes $\mathcal{O}(\zeta)$; while the latter does reduce to $\mathcal
O^{(0)}$ as $\zeta \to 0$, it is not obtained simply by promoting $g^{(0)} \to g(\zeta)$. Indeed, while there are corrections to $\mathcal{O}^{(0)}$ coming from the fact that the metric is no longer of the Kerr type, and not even type D more broadly, there are also corrections that can come from derivative couplings of the metric with the extra field, if the latter's configuration is nontrivial in the background.

Evidently, in general $\mathcal{S}(\zeta)\mathcal{E}(\zeta) \neq \mathcal{O}(\zeta)\mathcal{T}(\zeta)$, and one recovers the Wald identity in the limit $\zeta \rightarrow 0$, assuming that the metric $g(\zeta)$ is continuously connected to the Kerr spacetime.
We can rewrite this statement as $\mathcal{S}(\zeta)\mathcal{E}(\zeta) - \mathcal{O}(\zeta)\mathcal{T}(\zeta) = X(\zeta)$, where $X(\zeta)$ captures the deviation from the Wald identity and is such that, by definition, $X^{(0)}=0$.

Let us now assume that $X(\zeta)$ is known, and consider:
\begin{equation}
    \mathcal{S}(\zeta)\mathcal{E}(\zeta)h(\zeta) - \mathcal{O}(\zeta)\mathcal{T}(\zeta)h(\zeta) = X(\zeta)h(\zeta) \; .
\end{equation}
By definition, $\mathcal{T}(\zeta)h(\zeta)=\psi(\zeta)$, and hence we get:
\begin{equation}
      \mathcal{S}(\zeta)\mathcal{E}(\zeta)h(\zeta) - \mathcal{O}(\zeta)\psi(\zeta) = X(\zeta)h(\zeta) \; .
\end{equation}
Expanding in $\zeta$, neglecting $O(\zeta^2)$ terms, and using that $\mathcal{E}^{(0)}h^{(0)}=0$:
\begin{align}
    \mathcal{O}^{(0)} \psi^{(1)}  = & -\mathcal{O}^{(1)}\psi^{(0)} + \mathcal{S}^{(0)} (\mathcal{E}^{(0)}[h^{(1)}] +\mathcal{E}^{(1)}[h^{(0)}]) \notag \\
    &- X^{(1)} h^{(0)} \; .
\end{align}
Exploiting the $O(\zeta)$ term of the linearized Einstein equation \eqref{eq: Linearised Einstein equation bGR}, we arrive at the following modified Teukolsky equation
\begin{equation}
    \label{X generated Modified Teukolsky Equation}
    \mathcal{O}^{(0)} \psi^{(1)} = -\mathcal{O}^{(1)}\psi^{(0)} - X^{(1)}h^{(0)} + I_A^{(0)}[\varphi^{(1)}] + I_B^{(1)}[h^{(0)}]
\end{equation}
where:
\begin{align}
   I_A^{(0)}[\varphi^{(1)}] & = \mathcal{S}^{(0)} ( \tau^{(0)}_{\varphi}[\varphi^{(1)},g^{(0)}]) \; , \\
    I_B^{(1)}[h^{(0)}] & = \mathcal{S}^{(0)}(\tau^{(0)}_h[\Phi^{(1)},h^{(0)}]) \; .
\end{align}

In absence of an extra field, \eqref{X generated Modified Teukolsky Equation} becomes formally identical to  $\mathcal{O}^{(0)}\psi^{(0)}=0$ provided that the limit to GR as $\zeta\to 0$ is continuous. Indeed, $I_A, I_B$ would vanish due to the properties of $\tau_{\varphi}, \tau_h$, while the background spacetime reduces to the Kerr metric. As a consequence, $ X^{(1)}$ and $\mathcal{O}^{(1)}$ would vanish as well.  All contributions on the right-hand side of \eqref{X generated Modified Teukolsky Equation}, except of $I_A$, relate to changes in the background configuration of the metric, or the extra field. $I_A$ instead depends on the dynamical perturbation of the extra field. This is our main focus and an important feature of the modified Teukolsky equation, which we want to highlight.
As explained in detail in Appendix \ref{sec: appendix GHP}, the operator $\mathcal{S}^{(0)}$ is a GHP covariant quantity with GHP weights $p=4,q=0$, corresponding to a spin weight $s=(p-q)/2=2$. Hence, since the GHP weight of the extra field contributions $\tau_\varphi^{(0)}[\varphi^{(1)},g^{(0)}]$ is $p=0,q=0$, $I_A^{(0)}[\varphi^{(1)}]$  has spin weight $s=2$. This makes clear how a wave of an extra scalar field can be directly imprinted onto gravitational radiation.

It is worth stressing that our approach provides an intuitive understanding of the  structure and various terms of \eqref{X generated Modified Teukolsky Equation}, but it does not lend itself to directly computing these terms, except for $I_A^{(0)}[\varphi^{(1)}]$. A complete and explicit derivation has been given in  \cite{Li:2022pcy}. 
In \cref{app: MTE comparison}, we show explicitly that in the class of bGR theories we are considering as an example here, namely theories with extra scalar fields,
$I_A^{(0)}[\varphi^{(1)}]$ matches the source terms depending on the dynamical perturbation of the extra field that were obtained in \cite{Li:2022pcy}, 
 assuming  $\Phi^{(0)} = \varphi^{(0)} = 0$. Hence, these source terms depending on the dynamical perturbation of the scalar field can be simply obtained by acting with the operator $\mathcal{S}^{(0)}$ on the linearized (in both $\epsilon$ and $\zeta$) stress-energy tensor $\tau^{(0)}_{\varphi}[\varphi^{(1)},g^{(0)}]$ appearing in \eqref{eq: Linearised Einstein equation bGR}.

Finally, concerning the modified Teukolsky equation for $s=-2$, it reads:
\begin{equation}
    \label{Y generated Modified Teukolsky Equation}
    \mathcal{O}^{\dagger(0)} \tilde\psi^{(1)} = -\mathcal{O}^{\dagger(1)}\tilde\psi^{(0)} - Y^{(1)}h^{(0)} + L_A^{(0)}[\varphi^{(1)}] + L_B^{(1)}[h^{(0)}] .
    \end{equation}
Here, $Y^{(1)}$ comes from $Y(\zeta) = \mathcal{P}(\zeta)\mathcal{E}(\zeta)-\mathcal{O}^{\dagger}(\zeta)\mathcal{Q}(\zeta)$, where $\mathcal{Q}(\zeta)$ is the operator such that $\tilde\psi(\zeta) = \mathcal{Q}(\zeta)h(\zeta)$, and $\mathcal{P}(\zeta)$ reduces to $\mathcal{P}^{(0)}$ for $\zeta=0$, the latter being the operator that allows one to find a decoupled equation for $\tilde\psi^{(0)}$ in Petrov type-D, and is therefore the equivalent of $\mathcal{S}^{(0)}$. Moreover, $\mathcal{O}^\dagger (\zeta)$ is defined analogously to $\mathcal{O} (\zeta)$. Finally, similarly to the $s=2$ case:
\begin{align}
   L_A^{(0)}[\varphi^{(1)}] & = \mathcal{P}^{(0)} ( \tau^{(0)}_{\varphi}[\varphi^{(1)},g^{(0)}])\\
    L_B^{(1)}[h^{(0)}] & = \mathcal{P}^{(0)}(\tau^{(0)}_h[\Phi^{(1)},h^{(0)}]) \; .
\end{align}

\subsection{Dynamics of coupled fields in scalar-tensor theories}

To fully determine the source of \eqref{X generated Modified Teukolsky Equation} we need $h^{(0)}, \Phi^{(1)}, \varphi^{(1)}$.
The solutions for $\Phi^{(1)}$ and $\varphi^{(1)}$ can be obtained from \eqref{eq: stationary scalar} and \eqref{eq: Varphi1 Eqn}. Note also that the source term in the equation for $\varphi^{(1)}$ depends on $h^{(0)}$, which needs to be computed separately. 
To do so, we can solve the background Teukolsky equation for $s=2$, $\mathcal{O}^{(0)}\psi^{(0)}=0$, and for $s=-2$, $\mathcal{O}^{(0)\dagger}\tilde \psi^{(0)} = 0$, and reconstruct the metric perturbation $h^{(0)}$ from $\psi^{(0)}, \tilde \psi ^{(0)}$ using metric reconstruction techniques \cite{PhysRevD.11.2042,PhysRevD.19.1641}. Consequently, the equations for $\psi^{(0)}, \tilde\psi^{(0)},\Phi^{(1)},\varphi^{(1)},\psi^{(1)}$ form a hierarchical system:
\begin{align}
    \mathcal{O}^{(0)}\psi^{(0)} & =0 \;, \quad \mathcal{O}^{(0)\dagger}\tilde \psi^{(0)} = 0 \; ; \\
    \hat{\mathcal{O}}^{(0)} \Phi^{(1)} & = -\Sigma_B^{(0)}[g^{(0)}] \; ; \\
    \hat{\mathcal{O}}^{(0)} \varphi^{(1)} & =   -\sigma^{(0)}[\Phi^{(1)}, h^{(0)}] \; ; \\
    \mathcal{O}^{(0)} \psi^{(1)} & = -\mathcal{O}^{(1)}\psi^{(0)} - X^{(1)}h^{(0)} + I_A^{(0)}[\varphi^{(1)}] + I_B^{(1)}[h^{(0)}].
\end{align}

We can rewrite the equations in a way that better captures the dynamical sources: 
\begin{align}
    \mathcal{O}^{(0)} \psi^{(0)} & = 0\ ;  \\
    \hat{\mathcal{O}}^{(0)} \varphi^{(1)} & = -J^{(1)}[h^{(0)}]\ ;\\
    \mathcal{O}^{(0)} \psi^{(1)} & =  I_C^{(1)}[h^{(0)}]+I^{(0)}_A[\varphi^{(1)}]
     \; .
\end{align}
where
\begin{align}
J^{(1)}[h^{(0)}]  & = \sigma^{(0)}[\Phi^{(1)},h^{(0)}] \; ;
 \\ 
    I^{(1)}_C[h^{(0)}]  & = - \mathcal{O}^{(1)}[\psi^{(0)}] - X^{(1)}h^{(0)}  + I_B^{(1)}[h^{(0)}] \; .
\end{align}

\section{Ringdown ansatz in bGR theories} \label{sec: ringdown ansatz}

\subsection{Top-down approach}
Consider a BH ringing down to an equilibrium configuration in one of the bGR theories introduced in \cref{sec: setup}. Then, a far-away observer will measure a gravitational perturbation that in a time window some time after the merger can be expressed as:
\begin{equation}
    \psi(\zeta) \sim \sum_{\nn} c_{\nn}(\zeta) \psi_{\nn}(\zeta),
\end{equation}
where we employ the shorthand notation $\mathbf{n}$ to denote the angular momentum number $\mathfrak{l}$, the azimuthal number $\mathfrak{m}$, and the overtone number $\mathfrak{n}$, the $c_{\nn}$ are constant coefficients,  known as excitation coefficients, that depend on the initial data, and $\psi_{\nn}$ are the characteristic QNMs of the final bGR spacetime. They quantify the excitations of the characteristic quasinormal modes  $\psi_{\nn}(\zeta)$ of the final bGR BH.

As we are considering bGR theories whose BH solutions are continuously connected to the Kerr spacetime as $\zeta \rightarrow 0$, and further assuming that the initial data can be expanded as a power series in the coupling parameter $\zeta$, we can write:
\begin{equation}
\label{eq:topdown waveform expansion}
    \psi(\zeta) \sim \sum_{\nn} c^{(0)}_{\nn} \psi^{(0)}_{\nn} + \zeta \sum_{\nn} \left(  c^{(1)}_{\nn} \psi^{(0)}_{\nn} +  c^{(0)}_{\nn} \psi^{(1)}_{\nn} \right) + O(\zeta^2).
\end{equation}
The first term, of order $O(\zeta^0)$, captures the pure GR contribution to the signal: $\psi_{\nn}^{(0)}$ are Kerr QNMs and the excitation coefficients $c_{\nn}^{(0)}$ depend on the initial data computed at $\zeta=0$.

Assuming that we are considering times after some time $t_0$ from which linear perturbation theory is valid, and also assuming that for $\zeta \ll 1$ the bGR ringdown produces a waveform that is slightly deformed with respect to the one that would have been produced in a GR ringdown, we can expand the initial data as follows:
\begin{equation}
    \psi(\zeta, t_0) = \psi^{(0)}(t_0) + \zeta \psi^{(1)}(t_0) + O(\zeta^2)\; .
\end{equation}
The coefficients $c_{\nn}^{(1)}$ in \eqref{eq:topdown waveform expansion} depend on the initial data $\psi^{(1)}(t_0)$.

Finally, $\psi_{\nn}^{(1)}$ captures the correction to the set of Kerr QNMs. In principle, we have to allow for three different kinds of corrections. First, a frequency shift, due to the fact that the spectrum of the final spacetime $\{\omega( M,a,\zeta) \}$ depends on the bGR nature of the BH. This produces the shift:
\begin{equation}
    e^{-i \omega_{\nn}^{(0)}t}u^{(0)}(x^i) \rightarrow e^{-i(\omega^{(0)}_{\nn} + \zeta \omega^{(1)}_{\nn})t} u^{(0)}(x^i) \; ,
\end{equation}
where $u^{(0)}(x^i)$ is the spatial part of a Kerr QNM. The second correction is the shift to the spatial mode function $u^{(0)}(x^i)$, namely
\begin{equation}
    e^{-i \omega_{\nn}^{(0)}t}u^{(0)}(x^i) \rightarrow e^{-i \omega_{\nn}^{(0)}t}(u^{(0)}(x^i)+ \zeta u^{(1)}(x^i)) \; .
\end{equation}
The third possible correction is the least obvious one, coming from the consideration that the spectrum of the final spacetime could also feature a new set of frequencies $\{ \hat\omega_{\nn}\}$.  This picture is not in contrast with the assumption of the final BH being continuously connected to the Kerr metric, assuming that the new set of QNMs associated to $\hat\omega_{\nn}$ vanish as $\zeta \rightarrow 0$. Hence, at order $O(\zeta)$, this third correction has the form $\sim \zeta e^{-i \hat\omega_{\nn} t} v^{(1)}_{\nn}(x^i)$, with $v^{(0)}_{\nn}(x^i) = 0$.

In the next subsection we show constructively how all these different terms emerge from the study of the modified Teukolsky equation introduced in \ref{sec:Teukolsky}. Moreover, we also shed light on the physical meaning of $\hat{\omega}$, and why this new set of frequencies naturally arises in the context of the modified Teukolsky equation.

\subsection{Constructive approach}

As discussed in Section \ref{sec:Teukolsky}, the relevant set of equations governing the linear ringdown in a plethora of bGR theories can be brought in the form:
\begin{align}
    \mathcal{O}^{(0)} \psi^{(0)} & = 0\ ; \label{Teukolsky gravitational GR} \\
    \hat{\mathcal{O}}^{(0)} \varphi^{(1)} & = -J^{(1)}[h^{(0)}]\ ; \label{Teukolsky modified scalar} \\
    \mathcal{O}^{(0)} \psi^{(1)} & =  I^{(1)}_C[h^{(0)}]+I^{(0)}_A[\varphi^{(1)}]
    \label{Teukolsky modified gravitational} \; .
\end{align}
Note that this is not a set of coupled equations but rather a hierarchical system: 
GR gravitational perturbations $h^{(0)}$ are the dynamical source of scalar waves $\varphi^{(1)}$, which in turn source $\psi^{(1)}$.

We stress here that all the source terms have to be linear in the relevant dynamical field ($h^{(0)}$ or $\varphi^{(1)}$) and its derivatives, by construction of our perturbative formalism. For this reason, we can always rewrite a source term $\chi$ that has a dynamical dependence of the form $\chi[f(x^i)e^{-i \omega_1 t}+g(x^i)e^{-i \omega_2 t}]$, for some spatial functions $f,g$, as $\bar{\chi}(x^i,\omega_1) e^{-i \omega_1 t} + \bar{\chi}(x^i,\omega_2) e^{-i \omega_2 t}$, where $\bar{\chi}(x^i,\omega_{1,2})$ are spatial functions resulting from the separation of the time dependence.

This set of equations comes with initial data: $\psi^{(0)}_{t_0}, \dot{\psi}^{(0)}_{t_0}$ for the Teukolsky equation in the Kerr spacetime, $\varphi^{(1)}_{t_0}, \dot{\varphi}^{(1)}_{t_0}$ for the scalar equation, and $\psi^{(1)}_{t_0}, \dot{\psi}^{(1)}_{t_0}$ for the $O(\zeta)$ modified gravitational Teukolsky equation. 
The solution to \eqref{Teukolsky gravitational GR} is given by the convolution of the initial data with the Green's function.
Writing the Green's function as a sum of the free propagation term, branch cut contribution and QNM contribution \cite{Leaver:1986gd}, we have:
\begin{equation}
\label{GW in GR - all contributions}
    \psi^{(0)} = \psi^{(0)}_\mathrm{F} + \psi^{(0)}_\mathrm{B} + \sum_{\nn} c^{(0)}_{\nn}(x^i) \psi_{\nn}^{(0)} \; ,
\end{equation}
where $\psi_{\nn}^{(0)}  \sim e^{-i\omega_\nn^{(0)}t}$.
Let us study how a single GR QNM propagates through the hierarchical system. In this scenario \eqref{Teukolsky modified scalar} reads:
\begin{equation}
\label{eq: Ohat0 varphi1 eqn}
\hat{\mathcal{O}}^{(0)} \varphi^{(1)} = -J^{(1)}[h^{(0)}_\nn] \; ,
\end{equation}
in which $h^{(0)}_\nn$ is obtained from $\psi^{(0)}_\nn$, $\psi^{*(0)}_{\nn}$ by metric reconstruction.
The homogeneous solution is, exploiting the same arguments leading to \eqref{GW in GR - all contributions},
    \begin{equation}
\label{homogeneous scalar wave - all contributions}
    \varphi^{(1)}_\mathrm{hom} = \varphi^{(1)}_\mathrm{hom,F} + \varphi^{(1)}_\mathrm{hom,B} + \sum_{\nn} d^{(1)}_{\nn}(x^i) \varphi_{\nn}^{(0)} \;,
\end{equation}
where $\varphi_{\nn}^{(0)}$ are scalar QNMs, representing the characteristic oscillations of the scalar degree of freedom of the bGR BH: $\varphi_{\nn}^{(0)} = e^{-i \hat\omega^{(0)}_{\nn}t} \hat u^{(0)}(x^i)$. The particular solution to \eqref{eq: Ohat0 varphi1 eqn} is the one sourced by the selected GR QNM $h_{\nn}^{(0)}$. As the driving frequency is not a characteristic frequency of the QNM spectrum of $\hat{\mathcal{O}}^{(0)}$, no resonances arise, and a good ansatz for the solution is simply $\varphi^{(1)}_{gc} = F^{(1)}[h^{(0)}_{\nn}]$, where $F$ does not alter the time dependence of $h^{(0)}_{\nn}$ (implying that $\varphi^{(1)}_{gc}$ maintains the same time dependence as $h^{(0)}_{\nn}$). Here the subscript $gc$ stands for \textit{gravitationally contaminated}, indicating the presence of the driving term with frequency $\omega^{(0)}_\nn$. If we now plug $\varphi^{(1)} = \varphi_{\mathrm{hom}}^{(1)} + \varphi_{gc}^{(1)}$ into \eqref{Teukolsky modified gravitational}, and exploit that $I_A^{(0)}[\varphi^{(1)}]$ is linear in its argument, we have:
\begin{align}
\label{equation for mode shift and scalar led QNMs}
    \mathcal{O}^{(0)} \psi^{(1)} & =  I^{(1)}_C[h^{(0)}_\nn]
     + I_A^{(0)}[\varphi^{(1)}] \notag \\
     & = I^{(1)}[h^{(0)}_\nn] + I_A^{(0)}[\varphi^{(1)}_\mathrm{hom}] \;.
\end{align}
Here, $ I^{(1)}[h^{(0)}_\nn] = I_C^{(1)}[h^{(0)}_\nn] + I_A^{(0)}[\varphi^{(1)}_{gc}]$.
First, we focus on the solution driven by $h^{(0)}_\nn$, hence called $\psi^{(1)}_{\nn}$. Note that, in this instance, the driving frequency is a characteristic frequency of the QNM spectrum of $\mathcal{O}^{(0)}$. A good ansatz is:
\begin{equation}
\label{mode shift}
    \psi^{(1)}_{\nn} = -i \omega^{(1)}_{\nn}t \psi^{(0)}_{\nn}  + u^{(1)}(x^i)e^{-i \omega_{\nn}^{(0)}t} \;,
\end{equation}
where the frequency shift $\omega_{\nn}^{(1)}$ can be computed without solving the equation by following a perturbative approach described in \cite{Hussain:2022ins, Li:2023ulk}. A detailed study of this particular solution has been carried out in \cite{Li:2023ulk}; \eqref{mode shift} is consistent with their approach, in the limit of small $\zeta|\omega^{(1)}_\nn|t$. We will return to this condition in \cref{sec:Discussions}.

Let us now turn our attention to the solution to \eqref{equation for mode shift and scalar led QNMs} driven by $I_A^{(0)}[\varphi_\mathrm{hom}^{(1)}]$. In particular, we take for $\varphi_\mathrm{hom}^{(1)}$ a single scalar QNM out of all the possible contributions in \eqref{homogeneous scalar wave - all contributions}. The equation then becomes: 
\begin{equation}
    \mathcal{O}^{(0)} \psi^{(1)} = \bar{I}^{(1)}_A(x^i,\hat\omega^{(0)}_{\nn}) e^{-i \hat\omega^{(0)}_{\nn}t},
\end{equation}
with solution
\begin{equation}
    \psi^{(1)}_{\nn,sc} = q(x^i)e^{-i \hat\omega^{(0)}_{\nn}t},
\end{equation}
for some spatial function $q(x^i)$. As this solution arises from the driving scalar QNM, we call $\psi^{(1)}_{\nn,sc}$ \textit{scalar-contaminated} gravitational perturbations. Clearly, \eqref{Teukolsky modified gravitational} also has a homogeneous solution, of the form:
\begin{equation}
\label{GW in bGR - all contributions}
    \psi^{(1)}_\mathrm{hom} = \psi^{(1)}_\mathrm{F,hom} + \psi^{(1)}_\mathrm{B,hom} + \sum_{\nn} c^{(1)}_{\nn}(x^i) \psi_{\nn}^{(0)}.
\end{equation}

We now summarize the main results obtained. A GR QNM gets shifted as $\psi^{(0)}_{\nn} \rightarrow \psi^{(0)}_{\nn} + \zeta \psi^{(1)}_{\nn}$, and features a frequency shift term $\omega^{(1)}_{\nn}$ and a shifted spatial wavefunction $u^{(1)}_\nn(x^i)$. On the other hand, the excitation of a scalar QNM during the ringdown process leads to a scalar-contaminated order $O(\zeta)$ correction, of the form $\psi^{(1)}_{\nn,sc} \sim e^{-i \hat \omega^{(0)}_{\nn}t}$, to the gravitational perturbation. The gravitational initial data fix the coefficients $c^{(0)}_{\nn}(x^i), c^{(1)}_{\nn}(x^i)$ of the QNM driven evolution. Finally, we can write the QNM contribution to the bGR ringdown waveform as:
\begin{widetext}
\begin{equation}
    \label{ringdown waveform constructive approach}
    \psi_{\text{QNM}}(\zeta) = \sum_{\nn} c^{(0)}_{\nn}  \psi^{(0)}_{\nn} 
    + \zeta \sum_{\nn} \left( c^{(1)}_{\nn} \psi^{(0)}_{\mathbf{n}} + c_\nn^{(0)}\psi^{(1)}_\nn + c_{\nn}^{(0)}\psi_{\nn,sc}^{(1)}\right) + O(\zeta^2) \; .
\end{equation}
\end{widetext}

\subsection{Ringdown ansatz \label{sec:RingdownAnsatz}}
Having derived the QNM contribution to the bGR ringdown waveform, we analyze it near future null infinity at a fixed spatial coordinate. 
In this regime,
\eqref{ringdown waveform constructive approach} becomes:
\begin{widetext}
\begin{equation}
    \label{ringdown waveform observed far-away}
    \psi_{\text{QNM}}(\zeta) = \sum_{\nn} A_{\nn} e^{-i \omega^{(0)}_{\nn}t} + \zeta \sum_{\nn} \left( B_{\nn} e^{-i \omega^{(0)}_{\nn}t} -i A_{\nn}  \omega^{(1)}_{\nn}t e^{-i \omega^{(0)}_{\nn}t} + C_{\nn} e^{-i \hat\omega^{(0)}_{\nn}t}\right) + O(\zeta^2) \;.
\end{equation}
The linear growth in time can be understood as the first order expansion in $\zeta$ of $e^{-i(\omega_{\nn}^{(0)}+\zeta \omega_{\nn}^{(1)})t}$. Moreover, by only introducing an error at $O(\zeta^2)$ and defining $\tilde{A}_{\nn} \equiv A_{\nn}+\zeta B_{\nn}$, it is possible to rewrite \eqref{ringdown waveform observed far-away} as:
\begin{equation}
    \label{ringdown waveform observed far-away 2}
    \psi_{\text{QNM}}(\zeta) = \sum_{\nn} \tilde{A}_{\nn} e^{-i (\omega^{(0)}_{\nn} + \zeta \omega_{\nn}^{(1)})t} + \zeta \sum_{\nn}  C_{\nn} e^{-i \hat\omega^{(0)}_{\nn}t} + O(\zeta^2) \;.
\end{equation}
\end{widetext}
The two bGR corrections appear to be different in nature: the one featuring the frequency shift is a phase correction to the GR waveform, whereas the scalar-contaminated one looks more like an amplitude correction. We will return to this subtle point shortly.
The expression that we have derived here from first principles has the same structure as the ringdown ansatz recently proposed in \cite{Crescimbeni:2024sam} in the context of a theory-agnostic search. 

\section{Discussions \label{sec:Discussions}}

Our analysis is perturbative in $\zeta$, which is its major limitation. However, since bGR deviations have not been observed, it is reasonable to expect that $\zeta$ will indeed be small. For example, in sGB theories, recent constraints on the coupling constant for dilatonic ($f(\Phi) = e^\Phi$) theories are $\sqrt{\alpha_{\mathrm{sGB}}} \leq 0.31 \; \mathrm{km}$ \cite{Julie:2024fwy}.
This implies a dimensionless coupling $\zeta_\mathrm{sGB} \lesssim O(10^{-3})$ for BH masses $M \gtrsim 1 \; \mathrm{M}_\odot$.
On the other hand, using the METRICS formalism \cite{Chung:2023wkd,Chung:2023zdq,Chung:2024ira,Chung:2024vaf}, prospective constraints on dCS gravity were obtained in  \cite{Chung:2025gyg} by studying the frequency shift of the ringdown signal. These suggest that $\zeta_{\text{dCS}}\lesssim 0.1$ for $M \sim 10 \; \mathrm{M}_\odot$ and spin paramemeter  $\sim 0.7$.
Note, however, that by increasing the mass of the targeted remnant, one will inevitably find lower values of $\zeta_{\text{dCS}}$.
These are obviously  theory-dependent estimates and there could be theories which fall within the parameterisation \eqref{eq:ModifiedTheories} and are only weakly constrained, so that $\zeta$ takes high enough values that the ansatz \eqref{ringdown waveform observed far-away 2} is not sufficient.  We stress however that, although we resorted to perturbative treatment in $\zeta$ to demonstrate the frequency contamination in the ringdown, its physical origin is the coupling between the linear perturbations of the metric and the extra field, which is present irrespective of the size of $\zeta$ and indeed gets stronger (and more nuanced) as $\zeta$ increases. We therefore consider it a generic feature.

Assuming that $\zeta$ is indeed small, one can further simplify the ansatz \eqref{ringdown waveform observed far-away 2}.
Usually, the ringdown of a BH lasts $3-5$ damping times \cite{Carullo:2019flw} $\tau \sim 1/\mathrm{Im}(\omega_{220}^{(0)})$, which translates to a ringdown time $t_{R} \sim (30-80)M$. In what follows, we conservatively assume $t_R \sim O(10^2)$.
As $\mathrm{Re}(\omega^{(0)}_{220}) \sim O(10^{-1}) , \ \mathrm{Im}(\omega^{(0)}_{220}) \lesssim O(10^{-1})$, a conservative estimate is that the first order corrections to the frequencies are $\mathrm{Re}(\omega^{(1)}_{220} )\sim O(10^{-1}), \ \mathrm{Im}(\omega^{(1)}_{220} )\sim O(10^{-1})$, if the spectra of bGR theories are expected to be stable. 
Indeed, such an estimate is valid in the context of dCS and sGB, where $\omega^{(1)}_{220}$ has been computed \cite{Chung:2024vaf,Chung:2025gyg,Li:2025fci}. The above remarks imply that $\zeta |\omega_{220}^{(1)}| t_R \ll 1$ if $\zeta \lesssim O(10^{-2})$. Hence, considering the entire ringdown observation window, one has that for the first few dominant modes: $\zeta |\omega_{\nn}^{(1)}| t \ll 1$.
Whenever that is the case, the ansatz \eqref{ringdown waveform observed far-away 2} can be expanded as:
\begin{widetext}
    \begin{equation}
         \label{ringdown waveform observed far-away 4}
    \psi_{\text{QNM}}(\zeta) = \sum_{\nn} \tilde{A}_{\nn} e^{-i \omega^{(0)}_{\nn}t}\left(1 + \zeta \left( \frac{C_{\nn}}{\tilde{A}_{\nn}} e^{-i \Delta\omega^{(0)}_{\nn}t} -i \omega_{\nn}^{(1)}t  \right)  +O(\zeta^2) \right) \;,
    \end{equation}
\end{widetext}
where $\Delta\omega^{(0)}_\nn = \hat\omega^{(0)}_{\nn}-\omega^{(0)}_{\nn}$. Note that in this instance, the frequency shift and the modes sourced by extra fields are actually both $O(\zeta)$ time-dependent amplitude corrections to the GR waveform $\sum_{\nn}\tilde{A}_{\nn} e^{-i \omega^{(0)}_{\nn}t}$. Conversely, one could treat the two bGR corrections as modifications to the phase of the QNM. Indeed, starting from \eqref{ringdown waveform observed far-away 2}, one could reabsorb the modes with frequencies $\hat\omega^{(0)}_{\nn}$ into a second, time-dependent phase shift of the zeroth order QNM. To first order in $\zeta$, this is equivalent to \eqref{ringdown waveform observed far-away 4}. This is interesting to note, as phases are commonly considered to be easier to detect than amplitudes; if the condition that $\zeta |(\omega_{\nn}^{(1)}| t \ll 1$ holds, these two pictures are actually interchangeable.

We now compare the frequency shift correction and the scalar-contaminated one, starting from \eqref{ringdown waveform observed far-away 4}, and we show that, in general, the latter cannot be neglected. As a proof of principle, consider the following argument. If $\zeta \sim O(10^{-3})$, and assuming conservatively $\mathrm{Im}(\omega_{220}^{(1)})t_R  \sim O(10), \; \mathrm{Re}(\omega_{220}^{(1)})t_R  \sim O(10)$ we have $ \zeta \mathrm{Im}(\omega_{220}^{(1)})t_R \sim O(10^{-2}), \; \zeta \mathrm{Re}(\omega_{220}^{(1)})t_R \sim O(10^{-2})$, a value that justifies an expansion of $e^{-i (\omega^{(0)}_{\nn} + \zeta \omega_{\nn}^{(1)})t}$ up to first order in $\zeta$.  Using the known values of the frequencies of the (220) Kerr QNM ($\omega_{220}^{(0)}$) and of a scalar (220) QNM for a Kerr background ($\hat\omega_{220}^{(0)}$), we have that $\text{Im}(\hat\omega_{220}^{(0)}-\omega_{220}^{(0)}) \cdot t_R \sim O(1)$, and therefore $e^{ \mathrm{Im}(\hat\omega_{220}^{(0)}-\omega_{220}^{(0)})t} \sim 1$ for $t \in [0, t_R]$:
\begin{equation}
    \zeta \frac{C_{220}}{\tilde{A}_{220}} e^{\text{Im} (\hat\omega_{220}^{(0)}-\omega_{220}^{(0)})t_R} \sim \frac{C_{220}}{\tilde{A}_{220}} O(10^{-3}) \; .
\end{equation}
Comparing the real parts of the two bGR corrections, we now highlight the main implications of these remarks:
\begin{enumerate}[\itshape(i)]
\item  In the above, we have been very conservative in assuming $t_R \sim O(10^2)$. Comparing the two terms in the window $(20-30) M$, and assuming $\frac{C_{220}}{\tilde{A}_{220}} \sim O(1)$, one finds $\mathrm{Im}(\omega^{(1)}_{220})t/e^{\text{Im} (\hat\omega_{220}^{(0)}-\omega_{220}^{(0)})t} \in (2.3,3.8)$ for $\mathrm{Im}(\omega^{(1)}_{220}) = 0.1/M$. Hence the two corrections have comparable size.
\item 
Assuming that $\frac{C_{220}}{\tilde{A}_{220}} \sim O(1)$, the frequency shift term dominates in the late ringdown, near $t_R$. Conversely,
in the early ringdown, where the amplitudes are large and hence the signal is stronger, the frequency shift correction is comparable (as seen in $\textit{(i)}$) with the term causing the frequency contamination. However, as the ratio of the amplitudes depends on the physics of the merger and could in principle become much larger than $O(1)$ if nonlinearities source the new field during the merger (see \cite{DAddario:2023erc} for a discussion). In such scenarios, the frequency contamination term would be the dominant contribution at all times.

\end{enumerate}
Overall, it is clear that neglecting the scalar-contaminated correction to the bGR waveform could lead to potential biases in ringdown tests of GR. 

As a final comment, we note that if $p > 0 $ in \eqref{eq:ModifiedTheories}, the dimensionless parameter $\zeta$ decreases as the BH mass $M$ increases. As the size of bGR corrections to the waveform is controlled by $\zeta$, waveforms produced by Laser Interferometer Space Antenna (LISA) ringdown targets are likely to exhibit negligible deviations from GR, considering that such deviations have not already been detected in LIGO-Virgo-KAGRA sources (see also \cite{DAddario:2023erc}). This suggests that extreme mass ratio inspirals will be a more promising avenue for testing GR with LISA, compared to massive BH ringdown \cite{Maselli:2020zgv,Maselli:2021men,Barsanti:2022vvl,Spiers:2023cva,Speri:2024qak}. On the other hand, this could make massive BH ringdowns a perfect laboratory for studying the nonlinear ringdown in GR. 

\section{Conclusions}

In this work, we revisited the modelling of BH ringdown bGR, going beyond the standard paradigm in which deviations are captured solely through shifts in QNM frequencies. We systematically analyzed the structure of the ringdown signal as derived from modified Teukolsky equations in generic extensions of GR, with particular focus on theories that introduce additional degrees of freedom such as scalar fields.

After analyzing the structure of the modified Teukolsky equation \cite{Li:2022pcy}, we constructed a generalized ansatz for the ringdown waveform \eqref{ringdown waveform observed far-away 2}, valid in the small bGR coupling regime. This waveform incorporates two qualitatively distinct classes of deviations: (i) frequency shifts of the GR QNMs, and (ii) new gravitational modes that are sourced by the QNMs of extra fields. While both corrections appear at first order in the dimensionless coupling parameter $\zeta$, they differ in their origin and in their behavior over time. Importantly, we showed that under realistic assumptions for the duration of the ringdown window and current constraints on $\zeta$, the second class of corrections --- whose effect we referred to as frequency contamination --- are comparable to the frequency shift deviations. 

This behavior has clear implications for GW data analysis, suggesting that a search pipeline sensitive only to frequency shifts may overlook or mischaracterize important bGR effects. Our results also have implications for bGR QNMs in the frequency domain, as they highlight the importance of using an ansatz with an additional set of frequencies for every extra field when attempting to study the ringdown in bGR theories.

In part of our analysis we have assumed for concreteness that the extra field is a scalar. However, our main conclusions do not depend on the nature of the field, see \cite{Tomizuka:2025dpy} for a recent example, but solely on the assumption that it couples to gravity nonminimally and in a way that linear perturbations of the metric and the new field do not fully decouple. Our analysis has also assumed for simplicity that the new field is massless. A mass term can have a significant effect in the behaviour of both the background and the perturbations. We plan to report on this in future work.

\section{Acknowledgements}

We are grateful to Stephen Green, Andrea Maselli, Laura Sberna, Farid Thaalba, and Nicolás Yunes for enlightening discussions. T.P.S. acknowledges partial support from
the STFC Consolidated Grants no.~ST/X000672/1 and
no.~ST/V005596/1.

\appendix

\section{NP \& GHP formalism}\label{sec: appendix GHP}

\subsection{Main quantities in the NP formalism}

The study of the NP formalism \cite{Newman:1961qr} can be approached as a type of tetrad formalism. Consider an algebraically general spacetime and define a null, orthogonal tetrad $l^{\mu},n^{\mu},m^{\mu},m^{*\mu}$ such as $l^{\mu}n_{\mu}=1, m^{\mu}m^*_{\mu}=-1$ and with all other contractions equal to zero. As it turns out, by contracting tensorial quantities with the legs of the tetrad, one can define scalar quantities that are crucial to the study of gravitational radiation. 
First, we introduce five complex scalars by contracting the Weyl tensor $C_{\alpha \beta \gamma \delta}$ with different combinations of the tetrad vectors. The resulting quantities are called Weyl scalars and read:
\begin{align}
   & \Psi_0 = -C_{\alpha \beta \gamma \delta} l^{\alpha}m^{\beta}l^{\gamma}m^{\delta} \; ;
    \\ 
   &  \Psi_1 = -C_{\alpha \beta \gamma \delta} l^{\alpha}n^{\beta}l^{\gamma}m^{\delta} \; ;
   \\ 
   &  \Psi_2 = -\frac{1}{2} C_{\alpha \beta \gamma \delta} ( l^{\alpha}n^{\beta}l^{\gamma}n^{\delta} + l^{\alpha}n^{\beta}m^{\gamma}m^{*\delta} ) \; ;
    \\ 
   &  \Psi_3 = -C_{\alpha \beta \gamma \delta} l^{\alpha}n^{\beta}l^{\gamma}m^{\delta} \; ;
   \\ 
   &  \Psi_4 = -C_{\alpha \beta \gamma \delta} n^{\alpha}m^{*\beta}n^{\gamma}m^{*\delta} \; .
\end{align}
Directional derivatives are also introduced. This can be simply done as follows:
\begin{align}
    D & =l^{\alpha} \partial_\alpha \; ; \quad &\Delta&=n^{\alpha} \partial_\alpha \; ;
    \\ 
    \delta & = m^{\alpha} \partial_\alpha \; ; \quad &\delta^*&=m^{*\alpha} \partial_\alpha \; .
\end{align}
Finally, twelve complex scalars, called spin coefficients, play the role of Christoffel symbols. They read:
\begin{align}
    \kappa & = l^{\mu} m^{\nu} \nabla_{\mu} l_{\nu} \; ; \quad &\rho &= m^{*\mu} m^{\nu} \nabla_{\mu} l_{\nu} \; ; \\
    \sigma & = m^{\mu} m^{\nu} \nabla_{\mu} l_{\nu} \; ; \quad &\tau &= n^{\mu} m^{\nu} \nabla_{\mu} l_{\nu} \; ; \\
    \nu & = -n^{\mu} m^{*\nu} \nabla_{\mu} n_{\nu} \; ; \quad &\mu &= -m^{\mu} m^{*\nu} \nabla_{\mu} n_{\nu} \; ; \\
    \lambda & = - m^{*\mu} m^{*\nu} \nabla_{\mu} n_{\nu} \; ; \quad &\pi  &= - l^{\mu} m^{*\nu} \nabla_{\mu} n_{\nu} \; ;
    \end{align}
    \begin{align}
    \beta & = \frac{1}{2} ( m^{\mu}n^{\nu}\nabla_{\mu}l_{\nu} - m^{\mu} m^{* \nu} \nabla_{\mu} m_{\nu}  ) \; ;
    \\
    \alpha & = \frac{1}{2} ( m^{*\mu}m^{\nu}\nabla_{\mu}m_{*\nu} - m^{*\mu} l^{* \nu} \nabla_{\mu} n_{\nu}  ) \; ;
    \\
    \varepsilon & = \frac{1}{2} ( l^{\mu}n^{\nu}\nabla_{\mu}l_{\nu} - l^{\mu} m^{* \nu} \nabla_{\mu} m_{\nu}  ) \; ;
    \\
    \gamma & = \frac{1}{2} ( n^{\mu}m^{\nu}\nabla_{\mu}m_{*\nu} - n^{\mu} l^{* \nu} \nabla_{\mu} n_{\nu}  ) \; .
\end{align}

\subsection{GHP formalism}

The Geroch--Held--Penrose (GHP) formalism \cite{Geroch:1973am} is a refinement of the Newman--Penrose (NP) \cite{Newman:1961qr} framework 
and prioritizes the use of quantities that have specific transformation rules under a change of tetrad.
This improves efficiency in calculations, particularly in spacetimes admitting preferred null directions, such as Petrov type D ones, and gives extra structure to equations written in terms of GHP quantities.
The essence of the formalism is the concept of GHP covariance. Indeed, while the NP formalism relies on explicit components and directional derivatives associated with a chosen null tetrad $(l^\mu, n^\mu, m^\mu, m^{*\mu})$, the GHP formalism emphasizes objects that transform covariantly under a restricted class of tetrad transformations---namely spin-boost and null rotations about $l^\mu$ and $n^\mu$.

A geometrical quantity $\eta$ is said to be GHP covariant if, under the combined action of spin-boosts and phase rotations:
\begin{equation}
\label{tetrad transformation}
    (l^{\mu}, n^{\mu}, m^{\mu}, m^{*\mu}) \rightarrow (\Lambda l^{\mu}, \Lambda^{-1} n^{\mu}, e^{i\Gamma} m^{\mu}, e^{-i\Gamma} m^{*\mu}),
\end{equation}
where $\Lambda,\Gamma$ are smooth real-valued functions of the spacetime coordinates, it transforms as \cite{Geroch:1973am}:
\begin{equation}
    \eta \rightarrow \lambda^p \lambda^{*q} \eta,
\end{equation}
where $\lambda^2 = \Lambda e^{i \Gamma}$,with $(p,q)$ denoting the GHP weights. We say that $\eta$ is a GHP covariant quantity with weights $\eta \doteq (p,q)$. These weights encapsulate both the spin and boost behavior of the quantity, where the spin weight is given by $s = (p - q)/2$, and the boost-weight by $b = (p+q)/2$. This transformation rule allows for a compact and manifestly covariant treatment of spin-weighted fields.
Note that, as this section is just a review of the GHP formalism in GR, we drop the superscripts on fields and on operators indicating quantities perturbed with respect to the bGR coupling parameter $\zeta$.

From \eqref{tetrad transformation} it is clear that: $l^{\mu} \doteq (1,1),\ n^{\mu}\doteq (-1,-1),\ m^{\mu} \doteq (1,-1),\ m^{*\mu} \doteq (-1,1)$. Note that GHP weights behave additively under multiplication, that is, given $\eta \doteq (p,q), \kappa \doteq (p',q')$, $\eta \kappa \doteq (p+p', q+q')$.

 The Weyl scalars $\Psi_0 \equiv -C_{\alpha \beta \gamma \delta} l^{\alpha}m^{\beta}l^{\gamma}m^{\delta} ,\ \Psi_4 \equiv -C_{\alpha \beta \gamma \delta} n^{\alpha}m^{*\beta}n^{\gamma}m^{*\delta}$, as well as their dynamical perturbations $\psi_0$ and $\psi_4$, naturally acquire well-defined GHP weights and, consequently, spin weight. Specifically, the perturbed Weyl scalar $\psi \equiv  \psi_0$, has spin-weight $s=2$, whereas $\tilde{\psi} \equiv \Psi_2^{-4/3} \psi_4$ has spin weight $s=-2$.

A central object in the GHP formalism is the GHP covariant derivative $\Theta$, defined as \cite{Aksteiner:2014zyp}:
\begin{equation}
    \Theta_\mu = \nabla_\mu - s m^{*{\nu}}\nabla_{\mu}m^{\mu} - b n^{\nu} \nabla_{\mu} l_{\nu}.
\end{equation}
By definition, this operator is GHP weight-preserving, that is, if $\eta \doteq (p,q)$, then $\Theta \eta \doteq (p,q)$. One can then define new directional derivatives:
\begin{align}
    \text{\TH} & =l^{\alpha} \Theta_\alpha \; ; \quad & \text{\TH}' & =n^{\alpha} \Theta_\alpha \; ;
    \\ 
    \mathdh & = m^{\alpha} \Theta_\alpha \; ; \quad & \mathdh &=m^{*\alpha} \Theta_\alpha \; .
\end{align}
The GHP formalism introduces the operation (') that acts on a tetrad-dependent quantity by performing the following transformations \cite{Geroch:1973am}:
\begin{align}
    (l^{\mu})' & = n^{\mu} \; ; \quad  & (n^\mu)' &= l^\mu \; ;
    \\
    (m^{\mu})' & = m^{*\mu} \; ; \quad  & (m^{*\mu})' &= m^\mu \; .
\end{align}
One can then rename part of the NP spin coefficients as \cite{Geroch:1973am}:
\begin{align}
\nu &= -\kappa', & \lambda &= -\sigma', & \mu &= -\rho', \\
\pi &= -\tau', & \alpha &= -\beta', & \gamma &= -\varepsilon' \;.\label{eq:NPSpinCoeffsRule}
\end{align}
For $\eta \doteq (p, q)$, in terms of spin coefficients and directional derivatives in the NP formalism it holds that \cite{Geroch:1973am}:
\begin{equation}
    \text{\TH}\eta = ( D - p \varepsilon - q  \varepsilon^* )\eta\label{eq:Thorn}\; ;
\end{equation}
\begin{equation}
    \mathdh \eta= ( \delta - p \beta + q \beta^{*\prime} )\eta\label{eq:Edth} \; ,
\end{equation}

In a vacuum type-D spacetime, $\psi$ and $\tilde{\psi}$ obey, respectively, the Teukolsky equation for $s=2$, $\mathcal{O}\psi=0$, and for $s=-2$, $\mathcal{O}^{\dagger}\tilde \psi =0$. Where, the Teukolsky operator for $s=2$, $\mathcal{O}$, and for $s=-2$, $\mathcal{O}^{\dagger}$, can be expressed in terms of the GHP covariant derivative $\Theta$ as \cite{Aksteiner:2014zyp}:
\begin{align}
    & \mathcal{O}= g^{\mu \nu}(\Theta_{\mu} + 4 B_{\mu})(\Theta_{\nu}+4 B_{\nu}) - 16 \Psi_2 \ ; \\
    & \mathcal{O}^{\dagger}= g^{\mu \nu}(\Theta_{\mu} - 4 B_{\mu})(\Theta_{\nu}-4 B_{\nu}) - 16 \Psi_2,
\end{align}
where $\Psi_2$ is the background Weyl scalar with $s=0$, the only non-vanishing one in a Petrov type-D spacetime.
Note that we write the $s=-2$ Teukolsky operator as $\mathcal{O}^{\dagger}$ as it is the formal adjoint of $\mathcal{O}$ \cite{PhysRevLett.41.203}.

As explained in \ref{sec:Teukolsky}, the operator $\mathcal{O}$, together with the operators $\mathcal{S},\mathcal{E},\mathcal{T}$ satisfy:
\begin{equation}
    \mathcal{S}\mathcal{E} = \mathcal{O}\mathcal{T}.
\end{equation}
This operatorial identity is known as Wald identity. $\mathcal{E},\mathcal{S}, \mathcal{T}$ are defined in \eqref{linearized Einstein operator}, \eqref{S operator}, \eqref{T operator}.  From their definitions, and using that $l^{\mu}m^{\nu} \doteq (2,0)$, it follows that $\mathcal{S} \doteq \mathcal{T} \doteq (4,0)$, whereas $\mathcal{O} \doteq \mathcal{E} \doteq (0,0)$.  

\section{Comparison to the modified Teukolsky equation of Li et al. }\label{app: MTE comparison}

In this section, we compare the modified Teukolsky equation obtained from the deviation from the Wald identity with the one derived in \cite{Li:2022pcy}. In our notation, the latter reads:
\begin{equation}
\label{Li's equation}
    \mathcal{O}^{(0)}\psi^{(1)} = -\mathcal{O}^{(1)}\psi^{(0)} + K^{(1)}_{\mathrm{geo}}[h^{(0)}] + K^{(0)}_A[\varphi^{(1)}] + K_B^{(1)}[h^{(0)}],
\end{equation}
where $K^{(1)}_\mathrm{geo}[h^{(0)}]$ only depends on geometric quantities and hence is non-zero even in absence of extra-fields, whereas the operators $K^{(0)}_A,K^{(1)}_B$ vanish in absence of sources and extra fields, in the context of scalar-tensor theories, vanishing potential, and no higher derivative gravity modifications. To facilitate the comparison, note that we write their $\mathcal{S}^{(1,1)}_{\mathrm{geo}}$, appearing in their modified Teukolsky equation, eq. (95), as $\mathcal{S}^{(1,1)}_{\mathrm{geo}} = - \mathcal{O}^{(1)}\psi^{(0)}+K_{\mathrm{geo}}^{(1)}[h^{(0)}]$.

In this section we show explicitly that, in the class of theories analyzed in this study, namely scalar-tensor theories of gravity with vanishing potential $V(\Phi)$ and no higher derivative gravity corrections, one has:
\begin{equation}
\label{important check}
    I^{(0)}_A[\varphi^{(1)}] = \mathcal{S}^{(0)}(\tau_{\varphi}^{(0)}[\varphi^{(1)},g^{(0)}]) = K_A^{(0)}[\varphi^{(1)}] \; ,
\end{equation}
where $K_A^{(0)}[\varphi^{(1)}]$ has been introduced in \eqref{Li's equation}. 
In the original study \cite{Li:2022pcy}, a different notation is used. A quantity $Y^{(n,m)}$ is the $(\zeta=n,\epsilon=m)$ term of an expansion in $\zeta, \epsilon$ of a given object $Y$. In this section we use this notation whenever we deal with quantities defined in \cite{Li:2022pcy}. 
Using this notation, we rewrite: 
\begin{equation}
    K_A^{(0)}[\varphi^{(1)}]+K_B^{(1)}[h^{(0)}] \equiv K^{(1,1)},
    \label{eq:K11_V0}
\end{equation}
 where $K^{(1,1)}$ is the $S^{(1,1)}$ of eq. $(88)$ in \cite{Li:2022pcy}, and reads :
\begin{align}
   K^{(1,1)} & = \mathcal{E}_2^{(0,0)} S_2^{(1,1)} + \mathcal{E}_2^{(0,1)} S_2^{(1,0)} - \mathcal{E}_1^{(0,0)} S_1^{(1,1)} \notag\\
   & - \mathcal{E}_1^{(0,1)} S_1^{(1,0)} \; .
   \label{eq:K11}
\end{align}
Here:
 \begin{align}
S_1  = & (\delta + \pi^* - 2 \alpha^* - 2 \beta) \Phi_{00} - (D - 2 \varepsilon - 2 \rho^*) \Phi_{01} \notag\\
 & + 2 \sigma \Phi_{10} - 2 \kappa \Phi_{11} - \kappa^* \Phi_{02}  \label{eq:LiMathcalS1}\\
 S_2  = &(\delta + 2 \pi^* - 2 \beta) \Phi_{01} - (D - 2 \varepsilon + 2 \varepsilon^* - \rho^*) \Phi_{02} \notag \\
 & - \lambda^* \Phi_{00} + 2 \sigma \Phi_{11} - 2 \kappa \Phi_{12} \label{eq:LiMathcalS2} \\
    \mathcal{E}_1   = & \delta - \tau + \pi^* - \alpha^* - 3 \beta - \Psi_2^{-1} \delta \Psi_2 \label{eq:LiMathcalE1}\\
\quad
\mathcal{E}_2  = & D - \rho - \rho^* - 3 \varepsilon + \varepsilon^* - \Psi_2^{-1} D \Psi_2 \; ,
\label{eq:LiMathcalE2}
 \end{align}
where the expressions contain NP spin coefficients and derivative operators defined in \cref{sec: appendix GHP}, as well as the NP Ricci scalars
\begin{align}
\Phi_{00} &  = \frac{1}{2} R_{\mu \nu} l^\mu l^\nu = 4 \pi T_{ll} \\
\Phi_{01} & = \frac{1}{2} R_{\mu \nu} l^\mu m^\nu = 4\pi T_{lm} \\
\Phi_{10} & =  \frac{1}{2} R_{\mu \nu} l^\mu {m}^{*\nu} = 4 \pi T_{l m^*} \\
\Phi_{11} & =  \frac{1}{2} R_{\mu \nu} (l^\mu n^\nu + m^\mu {m}^{*\nu}) = 4\pi (T_{ln}+T_{m  m^*})\\
\Phi_{02} & = \frac{1}{2} R_{\mu \nu} m^\mu m^\nu = 4 \pi T_{m m} \\
\Phi_{12} & =  \frac{1}{2} R_{\mu \nu} n^\mu m^\nu = 4 \pi T_{nm}
\; .
\label{eq:NPRicciScalars}
\end{align}
Here, we employ the standard notation $T_{ll} = T_{\mu\nu}l^\mu l^\nu$ for projections onto the tetrad $\{ l^{\mu},n^{\mu},m^{\mu},m^{*\mu}\}$. As seen in \eqref{eq:K11_V0}, we aim to find the part of $K^{(1,1)}$ which contains $\varphi^{(1)}$ and show that is equivalent to $I_A^{(0)}[\varphi^{(1)}, g^{(0)}]$. 

We can now consider the perturbations of $\mathcal{E}_{1,2}$ and $S_{1,2}$ required for \eqref{eq:K11}. As seen from \eqref{eq:LiMathcalE1}-\eqref{eq:LiMathcalE2}, perturbations of $\mathcal{E}_{1,2}$ will not contain $\varphi^{(1)}$, and hence to find $K_A^{(0)}[\varphi^{(1)}]$ we will only have to consider the terms $\mathcal{E}_2^{(0,0)}S_2^{(1,1)}  -\mathcal{E}_1^{(0,0)}S_1^{(1,1)}$, and in particular contributions to $S_{1,2}^{(1,1)}$ from $T_{ij}^{(1,1)}$, where the Latin indices here are placeholders for the tetrad elements ${l,n,m,m^*}$.
For a Petrov type D background, one has that
\begin{equation}
    \sigma^{(0,0)} = \kappa^{(0,0)} = \lambda^{(0,0)} = 
    \nu^{(0,0)} =
    0
\end{equation}
\cite{Teukolsky:1973ha}.
Hence, one finds that for $S_{1,2}^{(1,1)}$, the only terms that survive from \eqref{eq:LiMathcalS1}-\eqref{eq:LiMathcalS2} and which contain $T^{(1,1)}_{ij}$ are
\begin{align}
    S_1^{(1,1)}  \supset & \; (\delta + \pi^* - 2 \alpha^* - 2 \beta)^{(0,0)} 4 \pi T_{ll}^{(1,1)} \notag \\ &
    - (D - 2 \varepsilon - 2 \rho^*)^{(0,0)} 4 \pi T_{lm}^{(1,1)} =S_{1A}^{(1,1)} \label{eq:S1_11v1}\\
     S_2^{(1,1)}  \supset & \; (\delta + 2 \pi^* - 2 \beta)^{(0,0)} 4 \pi T_{lm}^{(1,1)} \notag \\
     & - (D - 2 \varepsilon + 2 \varepsilon^* - \rho^*)^{(0,0)} 4 \pi T_{mm}^{(1,1)} =  S_{2A}^{(1,1)}\label{eq:S2_11v1}\; .
\end{align}
For $\mathcal{E}_{1,2}^{(0,0)}$, one finds that
\begin{align}
    \mathcal{E}_1^{(0,0)} & =  (\delta - 4\tau + \pi^* - \alpha^* - 3 \beta)^{(0,0)}\\
    \mathcal{E}_2^{(0,0)} & =  (D - 4\rho - \rho^* - 3 \varepsilon + \varepsilon^* )^{(0,0)} \; ,
\end{align}
where we have employed the vacuum type D relations
\begin{equation}
   (\Psi_2^{-1}\delta \Psi_2)^{(0,0)} = 3 \tau^{(0,0)} \; , \;  (\Psi_2^{-1}D \Psi_2)^{(0,0)} = 3 \rho ^{(0,0)} \; .
\end{equation}
We are therefore left with
\begin{widetext}
\begin{align}
    K^{(1,1)}  \supset & -  (\delta - 4\tau + \pi^* - \alpha^* - 3 \beta)^{(0,0)} \Big[(\delta + \pi^* - 2 \alpha^* - 2 \beta)^{(0,0)} 4 \pi T_{ll}^{(1,1)} 
    - (D - 2 \varepsilon - 2 \rho^*)^{(0,0)} 4 \pi T_{lm}^{(1,1)}\Big] 
    \notag \\ &
    +  (D - 4\rho - \rho^* - 3 \varepsilon + \varepsilon^*)^{(0,0)} \Big[(\delta + 2 \pi^* - 2 \beta)^{(0,0)} 4 \pi T_{lm}^{(1,1)}  - (D - 2 \varepsilon + 2 \varepsilon^* - \rho^*)^{(0,0)} 4 \pi T_{mm}^{(1,1)}\Big] \notag\\
    &  = K^{(1,1)}_A \; .
\label{eq:K11S}
\end{align}
\end{widetext}

The next step is to write this explicitly in the GHP formalism.
Considering once more $S_{1,2}^{(1,1)}$, we first observe that weights of the components of the stress-energy tensor are: 
\begin{equation}
    T^{(1,1)}_{ll}  \doteq (2,2) \; , \; T^{(1,1)}_{mm}  \doteq (2,-2) \; , \; T^{(1,1)}_{lm} \doteq (2,0) \; .
\end{equation}
Together with the definitions of $\text{\TH}$ and $\mathdh$, given in \eqref{eq:Thorn} and \eqref{eq:Edth} respectively, and with the spin coefficient relations \eqref{eq:NPSpinCoeffsRule}, this information implies that we can rewrite \eqref{eq:S1_11v1} and \eqref{eq:S2_11v1} as 
\begin{align}
    S_{1A}^{(1,1)} & = (\mathdh - \tau^{*\prime})^{(0,0)} 4 \pi T_{ll}^{(1,1)} 
    - (\text{\TH} - 2  \rho^*)^{(0,0)} 4 \pi T_{lm}^{(1,1)} \label{eq:S1_11v2}\\
    S_{2A}^{(1,1)} & = (\mathdh - 2  \tau^{*\prime})^{(0,0)} 4 \pi T_{lm}^{(1,1)} -(\text{\TH} -  \rho^*)^{(0,0)} 4 \pi T_{mm}^{(1,1)} \; . \label{eq:S2_11v2}
\end{align}
Then, the GHP weights of $S_{1A,2A}^{(1,1)}$ are
\begin{equation}
  S_{1A}^{(1,1)} \doteq (3,1) \;,\;  S_{2A}^{(1,1)} \doteq (3,-1) \; .
\end{equation}
Subsequently, we can look at the terms $ \mathcal{E}_2^{(0,0)} S_{2A}^{(1,1)}$ and $  \mathcal{E}_1^{(0,0)} S_{1A}^{(1,1)}$, employing \eqref{eq:S1_11v2} and \eqref{eq:S2_11v2}. As we did before, we use the definitions \eqref{eq:Thorn} and \eqref{eq:Edth}, plus the spin coefficient relations \eqref{eq:NPSpinCoeffsRule}, to find that
\begin{align}
\mathcal{E}_{1}^{(0,0)}S_{1A}^{(1,1)} & =  (\delta - 4\tau + \pi^* - \alpha^* - 3 \beta)^{(0,0)}S_{1A}^{(1,1)} \notag\\
    & =  (\mathdh -  \tau^{*\prime} - 4 \tau)S_{1A}^{(1,1)} \label{eq:E1S1}\\  
    \mathcal{E}_{2}^{(0,0)}S_{2A}^{(1,1)} & =  (D - 4\rho - \rho^* - 3 \varepsilon + \varepsilon^*)^{(0,0)}S_{2A}^{(1,1)}\notag\\
    & =  (\text{\TH} - 4 \rho-   \rho^*)S_{2A}^{(1,1)}\label{eq:E2S2}\; .
\end{align}
Overall, we can then write \eqref{eq:K11S} as
\begin{align}
    K^{(1,1)}_A  = & 4 \pi \bigg\{(\text{\TH} - 4 \rho-   \rho^*)^{(0,0)}\Big[   (\mathdh - 2 \tau^{*\prime})^{(0,0)}  T_{lm}^{(1,1)} \notag \\
    & -(\text{\TH} -  \rho^*)^{(0,0)} T_{mm}^{(1,1)}\Big] - (\mathdh -  \tau^{*\prime} - 4 \tau)^{(0,0)}\Big[ \notag \\
    & (\mathdh -  \tau^{*\prime})^{(0,0)} T_{ll}^{(1,1)}
    - (\text{\TH} - 2 \rho^*)^{(0,0)}T_{lm}^{(1,1)}\Big] \bigg\} \; .
\end{align}

By comparing this form of $K^{(1,1)}_A$ with $4\pi T_0$ in eq. (2.13) of \cite{Teukolsky:1973ha}, we see that this expression is none other than the linearised in $\zeta$ and $\epsilon$ source of the Teukolsky equation for $\psi_0$, obtained by acting with $\mathcal{S}^{(0)}$ on the source of the linearised Einstein equations. In our notation, we recall that this is precisely $\mathcal{S}^{(0)}(\tau^{(0)}_\varphi[\varphi^{(1)},g^{(0)}])$. Hence, we have shown that
\begin{equation}
    K_A^{(1,1)} = K_A^{(0)}[\varphi^{(1)}] = I_A^{(0)}[\varphi^{(1)}] = \mathcal{S}^{(0)}(\tau_\varphi^{(0)}[\varphi^{(1)},g^{(0)}]) \; .
\end{equation}

\bibliography{Refs.bib}

\end{document}